\documentclass[
 aps,           
pra,            
twocolumn,      
letterpaper,    
showpacs,       
preprintnumbers,
amsmath,        
amssymb,        
floatfix, 10pt]{revtex4-2}
\usepackage{graphicx}    
\usepackage{color}       
\usepackage{bm}          
\usepackage{braket}      
\usepackage{dcolumn}     
\usepackage[caption=false]{subfig}
\usepackage[T1]{fontenc}  
\usepackage[utf8]{inputenc} 
\newcommand{\I}{\text{I}}
\newcommand{\II}{\text{II}}
\begin{document}
	\title{\textbf{Evolution of quantum imaginarity in black hole quantum atmosphere} } 
	\author{Ruopu Sun and Xiaofen Huang}
	\email{huangxf1206@163.com} 
	\affiliation{School of Mathematics and Statistics, Hainan Normal University, Haikou 571158, China}

\begin{abstract}
Quantum imaginarity, as a critical resource metric for quantifying intrinsic nonreal coherence encoded in quantum states, exhibits nontrivial evolutionary behaviors in curved spacetime backgrounds. This work focuses on bipartite Dirac field reduced states in the quantum atmosphere of a static Schwarzschild black hole, aiming to explore the modulation of three mainstream imaginarity measures by Hawking thermal radiation. We reveal that the relative-entropy imaginarity , the geometric imaginarity and robustness of imaginarity, the fully accessible state presents a valley-shaped radial profile with a local minimum inside the quantum atmosphere, while the cross-coupling and fully inaccessible states follow opposite peak-shaped trends. Further analysis demonstrates that the Hartle–Hawking constant significantly strengthens the redistribution effect of imaginarity across both regions, whereas an increase in the event horizon radius weakens such redistribution and flattens the extremum features. These findings offer a new perspective for decoding the information structure of black hole quantum atmospheres and the intrinsic quantum nature of Hawking radiation.
\end{abstract}

\maketitle
\section{INTRODUCTION}

Quantum imaginarity serves as a rigorous quantification framework for intrinsic non-real quantum resources encoded in complex density matrices, possessing essential theoretical significance and broad application prospects in relativistic quantum information theory\cite{zhang2025quantum, 
wu2021multipartite, wu2020quantumcoherencewstate, hickey2018quantifyingimaginarity, xuan2025quantumimaginarityspeedlimit, zhang2025easilycomputablemeasuregaussian, chen2025imaginarityquantumchannels, xu2025coherenceimaginarityquantumstates, wei2024nonlocaladvantagesquantumimaginarity, xu2024quantifyingimaginarityquantumstates, guo2025quantifyingimaginaritydistance}. Originating from the complex off-diagonal elements of density operators, this physical quantity characterizes the unitary-invariant imaginary coherence that cannot be eliminated via arbitrary unitary transformations, which precisely captures the residual imaginary quantum resource after completely stripping all real-valued coherence contributions from quantum systems. Distinct from quantum entanglement, which vanishes in separable states, quantum imaginarity can persist in general separable mixed states and exhibits remarkable robustness against environmental noise and decoherence. Such unique advantages render imaginarity a more flexible and reliable observable for exploring quantum information dynamics in curved spacetime backgrounds, especially for describing quantum correlation behaviors near black hole horizons. 

The fundamental necessity of complex quantum amplitudes has been theoretically argued and experimentally validated in recent works, confirming that real-valued quantum formulations are fundamentally insufficient to reconstruct all observable quantum phenomena\cite{Renou2021,Wu2024}. On this basis, Hickey and Gour established a complete and self-consistent resource-theoretic framework of quantum imaginarity, where real density matrices and real-preserving quantum operations are strictly defined as free states and free transformations, thereby formally establishing imaginarity as a fundamental and independent quantum resource beyond entanglement and coherence. To quantitatively characterize the magnitude of imaginary resources, a series of systematic metrics have been successively proposed, including trace-norm robustness, relative-entropy imaginarity, and geometric imaginarity, which are tailored for different operational tasks such as quantum state discrimination, quantum resource manipulation, and quantum information transmission\cite{bera2026strongnonlocalityimaginarityentanglement}
. With the rapid development of resource theory, contemporary studies have further explored nonlocal imaginarity characteristics in multipartite quantum systems, revealed the intricate interplay between imaginarity and conventional quantum resources, and clarified their conversion and complementarity mechanisms\cite{Li2023}. Furthermore, recent advances have extended the imaginarity paradigm from discrete-variable quantum systems to continuous-variable configurations and relativistic quantum scenarios, significantly enriching the theoretical connotation and expanding the applicable boundary of imaginary quantum resource theory\cite{Zhou2025}.

Hawking radiation laid the foundational bridge linking general relativity and quantum mechanics, marking a landmark breakthrough in modern theoretical physics\cite{barenboim2026quantumcorrectedbondimass2d, pcastro2025modifiedhawkingradiationquantum, li2024informationretrievalhawkingradiation, shi2023quantumsimulationhawkingcurvedsuperconducting, syu2022analogoushawkingentanglementbec, singh2017quantumgravityeffectshawking}. Unraveling the dynamic evolution of imaginary quantum resources and quantum correlations within curved spacetimes is therefore critical to resolving the long-standing black hole information paradox. The standard naive picture claims that Hawking radiation originates from quantum fluctuations confined to a narrow near-horizon strip satisfying $\Delta r=r-r_h\ll r_h$, with $r_h$ denoting the event horizon radius. Nevertheless, full tensor analyses of total emission rates associated with Hawking radiation demonstrate that thermal radiation is produced across a broader radial zone termed the black hole quantum atmosphere, where the radial range satisfies $\Delta r = r_A - r_h \sim r_h$\cite {giddings2006blackholeinformationunitarity,hod2016stefan,barbado2016tensor,ong2020rnatmosphere,dey2019freefallatmosphere,liu2026coherencecvatmosphere}. This updated understanding of the spatial extent of Hawking radiation motivates us to revisit the dynamic evolution of imaginary quantum resources over this extended region.

Recent progress in relativistic quantum information has concentrated on non-inertial curved spacetimes, with abundant studies exploring the variation of quantum correlations, including entanglement, quantum coherence\cite{wu2023coherencedesitter,wu2024curvatureenhancedmultipartitecoherencemultiverse}, quantum discord\cite{wang2010discordnoninertial}, and quantum steering\cite{du2024maximalsteered,wang2014discorddilaton,liao2025quantumcoherence}. Key results confirm that these quantum correlations undergo inevitable degradation induced by Hawking radiation effects, findings that deepen our comprehension of quantum information transport on curved gravitational backgrounds and the core puzzles of the black hole information paradox \cite {elghaayda2024accessiblecorr,zhang2025entanglement}. Despite substantial theoretical advances, the local evolution of quantum imaginarity within the black hole quantum atmosphere remains largely unstudied, and the systematic modulation of imaginary quantum resources by critical geometric and thermal parameters has not yet been fully clarified.

In this work, we study the radial evolution of the three different measures of quantum imaginarity in the quantum atmosphere of a static Schwarzschild black hole. Using position-dependent Hartle–Hawking local temperatures to characterize near-horizon thermal effects, we obtain closed-form formulas of the three measures for bipartite Dirac-field reduced states over accessible Region I and inaccessible Region II. Systematic numerical analysis indicates that $r_h$ and $D_{HH}$ oppositely modulate the redistribution of imaginary quantum resources. This work yields new insights into the internal information structure of black hole quantum atmospheres and the quantum origin of Hawking radiation.

The remainder of this paper is organized as follows. In Section II, we briefly introduce the vacuum structure of Dirac fields in Schwarzschild spacetime. In Section III, we construct a bipartite entangled pure state within the framework of the local Hawking effect and adopt the Hartle–Hawking temperature model. We sequentially investigate the evolutionary behaviors of reduced states belonging to physically accessible and inaccessible regions characterized by three imaginarity quantifiers, namely relative-entropy imaginarity, geometric imaginarity and robustness-based imaginarity. Finally, Section IV concludes the main findings of this work.

\section{Quantumness of Schwarzschild spacetime}

	The Schwarzschild solution is the simplest vacuum black hole of general relativity, describing a static, uncharged, spherically symmetric spacetime. The Schwarzschild metric reads
\begin{equation}
	\begin{split}
		ds^2 &= -\left(1-\frac{2M}{r}\right)dt^2 + \left(1-\frac{2M}{r}\right)^{-1}dr^2 \\
		&\quad + r^2d\theta^2 + r^2\sin^2\theta d\phi^2,
	\end{split}
	\label{eq:sch_metric}
\end{equation}
where $M$ denotes the black hole mass, $r$ is the radial coordinate, $t$ denotes the time coordinate, $\theta$ denotes the polar angle, and $\phi$ denotes the azimuthal angle. Also, we adopt natural units where $G = c = \hbar = k_B = 1$ for simplicity.
The Dirac equation in Schwarzschild spacetime reads
\begin{equation}
	\left[\gamma^\alpha e_\alpha^\mu\left(\partial_\mu + \Gamma_\mu\right)\right]\psi = 0,
\end{equation}
where $\gamma^\alpha$ denotes the Dirac matrices, $e^\mu_\alpha$ represents the vierbein, and $\Gamma_\mu$ stands for the spin connection coefficients. Substituting the Schwarzschild vierbein into the Dirac equation, the Dirac equation for Schwarzschild spacetime can be expressed as
\begin{equation}
	\begin{split}
		&\gamma_1\sqrt{1-\frac{2M}{r}}\left[\frac{\partial}{\partial r}+\frac{1}{r}+\frac{M}{2r(r-2M)}\right]\psi-\frac{\gamma_0}{\sqrt{1-\frac{2M}{r}}}\frac{\partial \psi}{\partial t}\\
		&+\frac{\gamma_3}{r\sin\theta}\frac{\partial \psi}{\partial \varphi}+\frac{\gamma_2}{r}\left(\frac{\partial}{\partial \theta}+\frac{\cot\theta}{2}\right)\psi =0,
	\end{split}
	\label{eq:dirac_full}
\end{equation}
where $\gamma_k$ ($k=0,1,2,3$) are the Dirac matrices. 
This equation admits a set of positive-frequency outgoing solutions (Schwarzschild modes) localized inside and outside the event horizon, which are explicitly written as
\begin{equation}
	\begin{cases}
		\psi_k^{\I+} = \varsigma(r)e^{-i\omega u},\\
		\psi_k^{\II+} = \varsigma(r)e^{i\omega u}.
	\end{cases}
	\label{eq:fermi_modes}
\end{equation}
where $\psi_k^{\I+}$ and $\psi_k^{\II+}$ correspond to the positive-frequency solutions outside (Region $\I$) and inside (Region $\II$) the event horizon, $\omega$ denotes the monochromatic frequency of the Dirac field, $\varsigma(r)$ represents the four-component Dirac spinor, and the degenerate coordinate $u$ is defined as $u = t - r_*$, where $r_* = r + 2M\ln\left(\frac{r}{2M}-1\right)$. Using the Damour--Ruffini method, the positive energy Kruskal modes bridge the two families of Schwarzschild modes in Eq.\eqref{eq:fermi_modes}:
\begin{equation}
	\begin{split}
		\Phi_{k,\I}^+ &= e^{-2\pi M\omega}\Psi_{-k,\II}^- + e^{2\pi M\omega}\Psi_{k,\I}^+, \\
		\Phi_{k,\II}^+ &= e^{-2\pi M\omega}\Psi_{-k,\I}^- + e^{2\pi M\omega}\Psi_{k,\II}^+.
	\end{split}
	\label{eq:bogo_kruskal}
\end{equation}
Under the appropriate Kruskal modes, the Dirac field can be expanded as
\begin{equation}
\begin{split}
\psi ={}& \int dk\,\big[2\cosh(4\pi M\omega_i)\big]^{-\frac12}
\Big[ \hat{c}_k^{\mathrm{II}}\Psi_{k,\mathrm{II}}^+ + \hat{d}_{-k}^{\mathrm{II}\dagger}\Psi_{-k,\mathrm{II}}^- \\
&+ \hat{c}_k^{\mathrm{I}}\Psi_{k,\mathrm{I}}^+ + \hat{d}_{-k}^{\mathrm{I}\dagger}\Psi_{-k,\mathrm{I}}^-\Big],
\end{split}
\label{eq:field_expansion}
\end{equation}
where the operators $\hat{c}_k$ and $\hat{d}_{-k}^\dagger$ serve as the creation and annihilation operators that operate on the Kruskal vacuum. Using the Bogoliubov transform, the Kruskal vacuum state and excited state in Schwarzschild spacetime are expressed as
\begin{equation}
	\begin{aligned}
		|0\rangle_k &= \frac{1}{\sqrt{e^{-\frac{\omega}{T}} + 1}}|0\rangle_\mathrm{I}|0\rangle_\mathrm{II} + \frac{1}{\sqrt{e^{\frac{\omega}{T}} + 1}}|1\rangle_\mathrm{I}|1\rangle_\mathrm{II}, \\
		|1\rangle_k &= |1\rangle_\mathrm{I}|0\rangle_\mathrm{II},
	\end{aligned}
	\label{eq:kruskal_vac}
\end{equation}
where $T = \frac{1}{8\pi M}$ denotes the Hawking temperature, $\omega$ is the particle frequency. To simplify the subsequent analysis, we define the substitution $\frac{1}{\sqrt{e^{-\frac{\omega}{T}} + 1}} = \cos r$ and $\frac{1}{\sqrt{e^{\frac{\omega}{T}} + 1}} = \sin r$, where $r$ denotes the parameter\cite {eune2019test, wu2024fermionicsteering}.

\section{QUANTUM IMAGINARITY IN BLACK HOLE QUANTUM ATMOSPHERE}
\subsection*{A. Relative-entropy imaginarity }
To characterize the imaginary quantum resources in curved spacetime backgrounds, we consider a bipartite entangled pure state. It provides a simple and tunable framework for analyzing how imaginary quantum resources evolve under gravitational spacetime effects. The state is given by
\begin{equation}
	|\gamma_{AB}\rangle = \cos\frac{\theta}{2}|00\rangle + e^{i\phi}\sin\frac{\theta}{2}|11\rangle,
	\label{eq:pure_gamma}
\end{equation}
where $\theta\in[0,\pi]$ and $\phi\in[0,2\pi]$\cite {han2026quantificationresourcetheoryimaginarity}.

Relative-entropy imaginarity originates from the operational resource theory of imaginarity \cite{Wu2021}. In this formalism, real density matrices constitute free states, and valid dynamical maps are restricted to real-preserving operations. The quantifier quantifies imaginary resources by contrasting a state with its real projection $\mathrm{Re}\rho$, obtained as the symmetric average of $\rho$ and its transpose. Its definition reads
\begin{equation}
\mathcal{I}_{r}(\rho)=S(\mathrm{Re}\,\rho)-S(\rho),
\label{eq:Ir_def}
\end{equation}
where $\mathrm{Re}\,\rho = \tfrac12(\rho+\rho^\mathsf{T})$, and $\rho^\mathsf{T}$ stands for matrix transposition \cite{li2026multistateimaginarity}.
This quantity carries a clear operational meaning: it gives the minimal thermodynamic cost to erase all imaginary resources using free real-preserving operations. While imaginarity is closely related to quantum coherence, it focuses solely on nonreal matrix elements and provides an independent viewpoint to quantify imaginary quantum resources.

In the bipartite pure state introduced above, subsystems A and B correspond to two observers, Alice and Bob, respectively. We suppose that Alice and Bob both hover outside the event horizon of a Schwarzschild black hole with identical radial coordinate parameter $r$. Owing to Hawking thermal radiation, the Dirac field mode decomposition undergoes distortion from the viewpoint of stationary observers near the horizon. Consequently, the initial bipartite state $\rho_{AB}$ evolves into a four-partite global state $\rho_{A_{\mathrm{I}}A_{\mathrm{II}}B_{\mathrm{I}}B_{\mathrm{II}}}$. After tracing out the unobservable interior modes $A_{\mathrm{II}}$ and $B_{\mathrm{II}}$, the physically accessible bipartite reduced state $\rho_{A_{\mathrm{I}}B_{\mathrm{I}}}$ is derived, which takes the explicit matrix form
\begin{equation}
\begin{split}
&\rho_{A_{\mathrm{I}}B_{\mathrm{I}}}=
 \\
&
\begin{pmatrix}
	c^2(1-t)^2 & 0 & 0 & e^{-i\phi}cs(1-t) \\
	0 & c^2 t(1-t) & 0 & 0 \\
	0 & 0 & c^2 t(1-t) & 0 \\
	e^{i\phi}cs(1-t) & 0 & 0 & c^2 t^2 + s^2
\end{pmatrix},
\end{split}
\label{eq:rho_A1B1_ultra}
\end{equation}
where $c=\cos\tfrac{\theta}{2}$, $s=\sin\tfrac{\theta}{2}$, and $t=\sin^2 r$ denote shorthand parameters for simplicity.

According to the definition of relative-entropy imaginarity in Eq. \eqref{eq:Ir_def}, the relative-entropy imaginarity of the reduced state $\rho_{A_{\mathrm{I}}B_{\mathrm{I}}}$ can be computed as follows:

\begin{align}
\begin{split}
	\mathcal{I}_r(\rho_{A_{\mathrm{I}}B_{\mathrm{I}}})
	&=\frac{1}{2\log 2}
	\Bigl[
	f(u_{11},v_{11},w_{11})
	-f(\overline{u}_{11},v_{11},w_{11})
	\Bigr],
	\label{eq:Ir_A1B1}\\[6pt]
    \end{split}
\end{align}
where 
\begin{align*}
	u_{11}&=4c_\theta^2c_r^4-4c_\theta^2c_r^2+1,\\
	\overline{u}_{11}&=(2c_\theta^2c_r^4+s_\theta^2)^2
	-4c_\theta^2c_r^4(s_\phi^2s_\theta^2-s_r^4s_\theta^2+s_r^4),\\
	v_{11}&=2c_\theta^2c_r^4+s_\theta^2,\\
	w_{11}&=c_\theta^2c_r^4+\tfrac{1}{2}s_\theta^2,\\[4pt]
\end{align*}
and the compact trigonometric shorthands are defined as $c_\theta=\cos\left(\frac{\theta}{2}\right)$, $s_\theta=\sin\left(\frac{\theta}{2}\right)$,
$c_r=\cos r$, $s_r=\sin r$, $c_\phi=\cos\phi$, and $s_\phi=\sin\phi$.

Notably, Eq. (\ref{eq:Ir_A1B1}) relies on a unified auxiliary function $f(u,v,w)$ encapsulating the logarithmic entropy difference, whose concrete form is given by
\begin{equation}
	\begin{split}
		f(u,v,w)
		&=\bigl(\sqrt{u}+v\bigr)\log\!\left(w+\frac{\sqrt{u}}{2}\right) \\
		&\quad-\bigl(\sqrt{u}-v\bigr)\log\!\left(w-\frac{\sqrt{u}}{2}\right).
	\end{split}
	\label{eq:L_function}
\end{equation}
Here $u$, $v$, $w$ are state-dependent auxiliary quantities constructed from the matrix entries of $\rho_{A_{\mathrm{I}}B_{\mathrm{I}}}$. 

Similarly, the relative-entropy imaginarity of the other reduced states $\rho_{A_{\mathrm{II}}B_{\mathrm{II}}}$, $\rho_{A_{\mathrm{I}}B_{\mathrm{II}}}$, and $\rho_{A_{\mathrm{II}}B_{\mathrm{I}}}$ can also be computed, and their explicit expressions are given in the Appendix.

\begin{figure*}[tb]
	\centering
	\subfloat[]{%
		\includegraphics[width=0.32\textwidth]{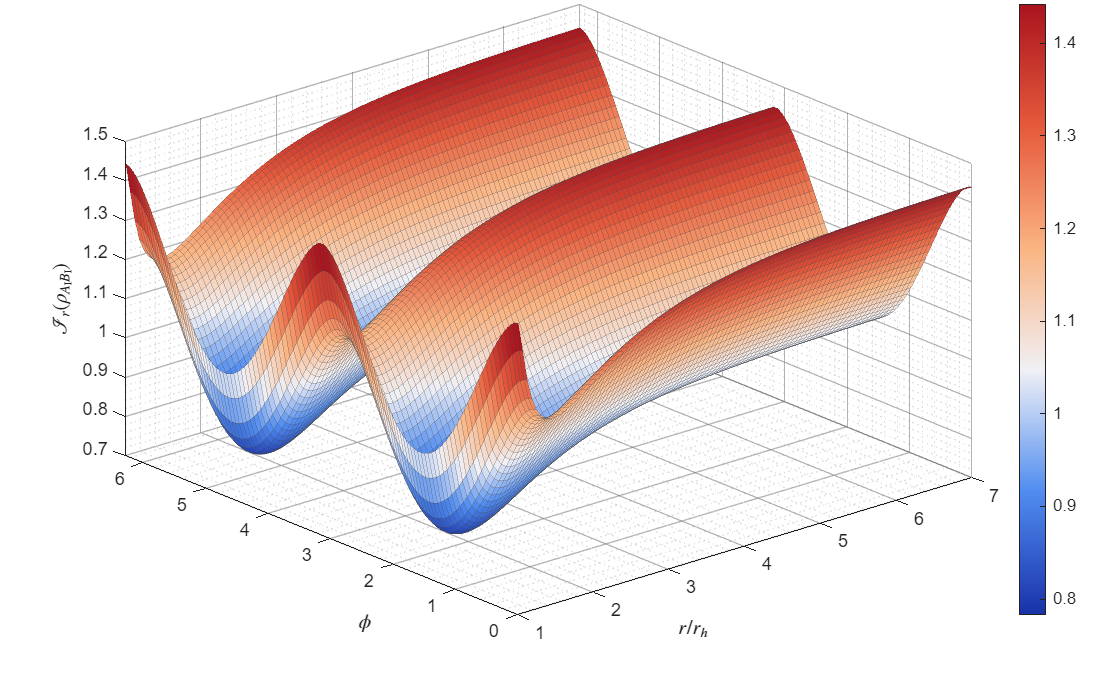}%
		\label{subfig:Ir_A1B1_curve}%
	}
	\hfill
	\subfloat[]{%
		\includegraphics[width=0.32\textwidth]{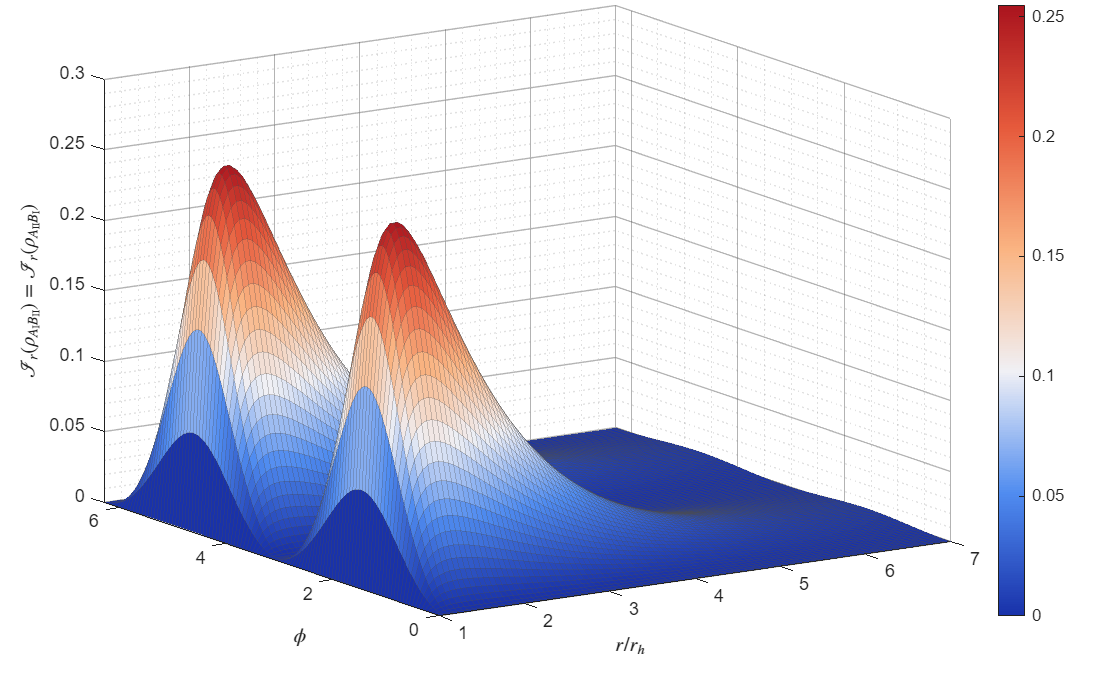}%
		\label{subfig:Ir_A1B2_curve}%
	}
	\hfill
	\subfloat[]{%
		\includegraphics[width=0.32\textwidth]{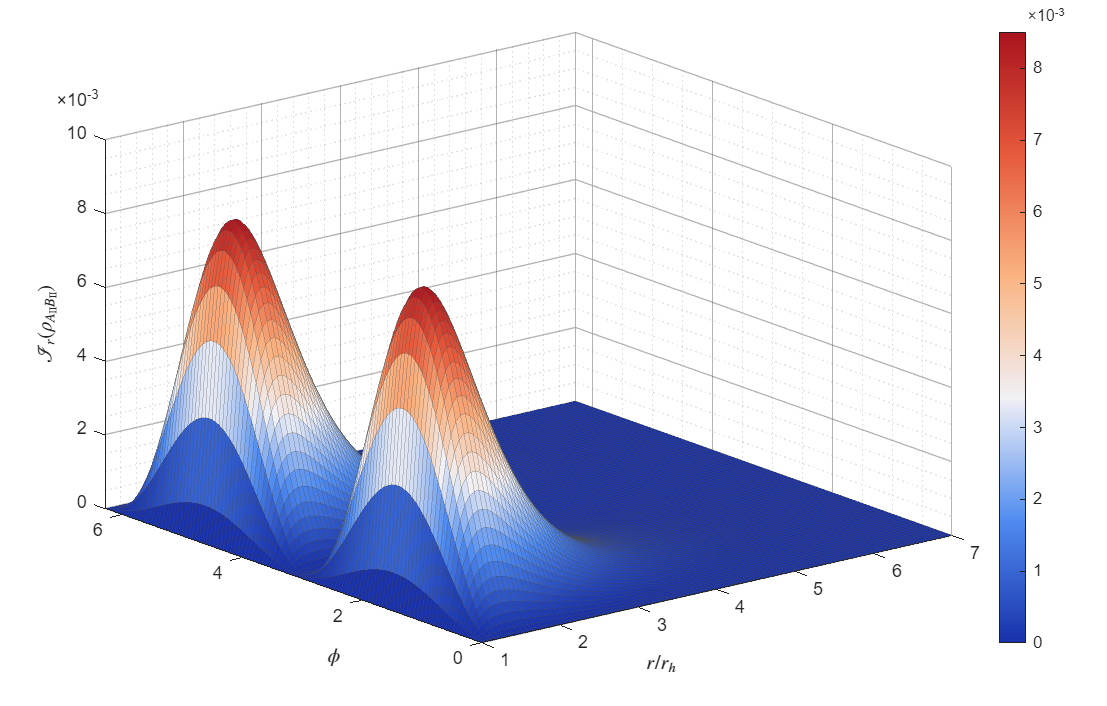}%
		\label{subfig:Ir_A2B2_curve}%
	}

	\vspace{0.6em} 

	\subfloat[]{%
		\includegraphics[width=0.32\textwidth]{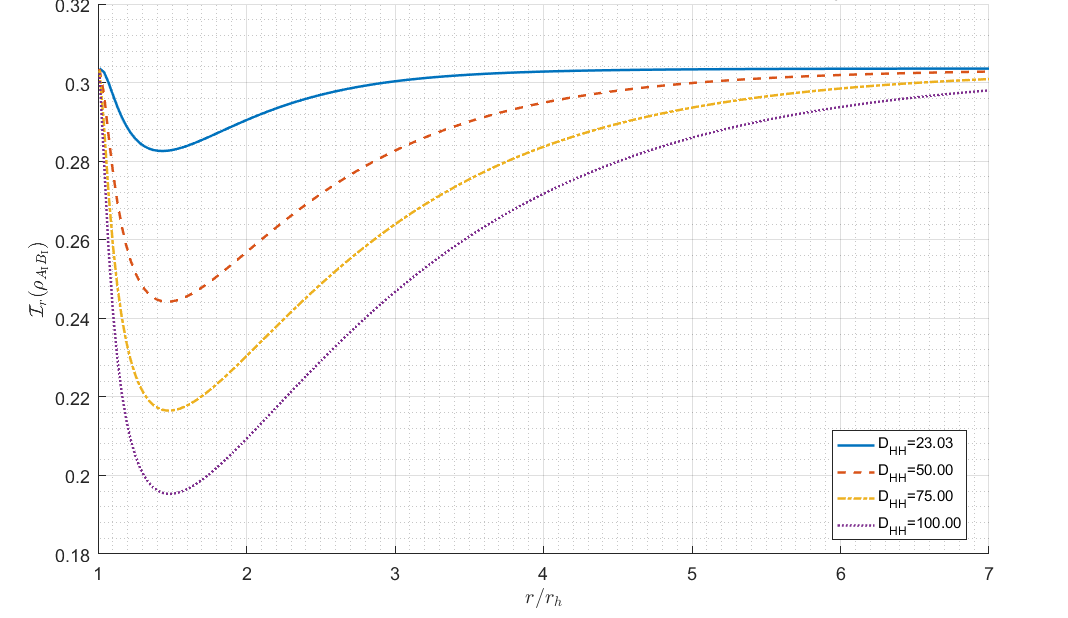}%
		\label{subfig:Ir_A1B1_surf}%
	}
	\hfill
	\subfloat[]{%
		\includegraphics[width=0.32\textwidth]{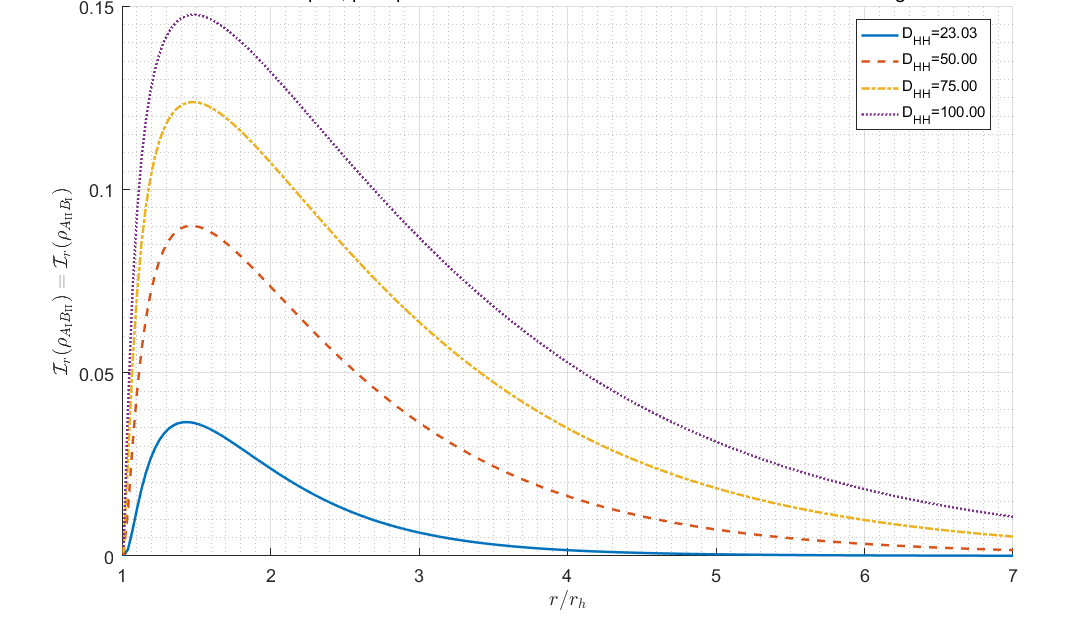}%
		\label{subfig:Ir_A1B2_surf}%
	}
	\hfill
	\subfloat[]{%
		\includegraphics[width=0.32\textwidth]{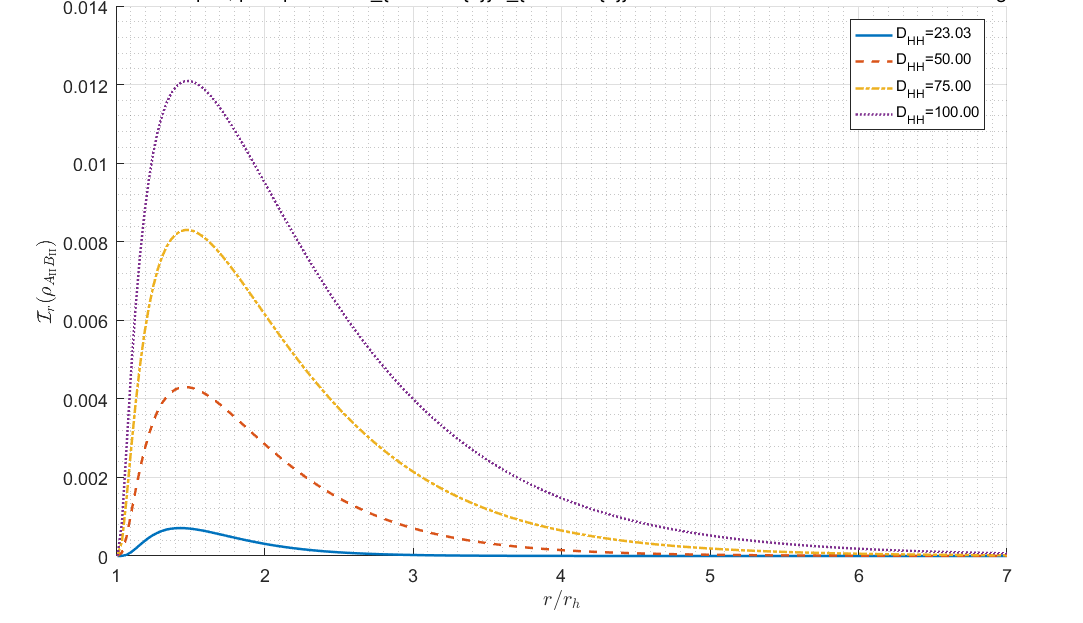}%
		\label{subfig:Ir_A2B2_surf}%
	}

	\caption{Panels (a)--(c) display three‑dimensional surface plots of imaginarity as functions of normalized radial distance $r/r_h$ and relative phase $\phi$, with the state parameter fixed at $\theta=\pi/4$, $D_{HH}=50$, and field frequency $\omega=1$. Panels (d)--(f) show the corresponding curves of imaginarity varying with $r/r_h$ under fixed $\theta$ and $\phi$.}
	\label{fig:Ir_all_3D}
\end{figure*}

To describe the the effect of the black hole atmosphere, we adopt the local temperature $T_{HH}$ defined under the Hartle–Hawking vacuum background to replace the global Hawking temperature $T$, which can be expressed as
\begin{equation}
	\begin{aligned}
		T_{HH} &= T_H \sqrt{1-\frac{r_h}{r}} \\
		&\quad \cdot \sqrt{1+2\frac{r_h}{r}+\left(\frac{r_h}{r}\right)^2\left(9+4D_{HH}+36\ln\left(\frac{r_h}{r}\right)\right)},
	\end{aligned}
	\label{eq:THH}
\end{equation}
where $T_H = 1/(4\pi r_h)$, and $D_{HH}$ represents the constant in the stress tensor associated with the Hartle-Hawking vacuum. The constant $D_{HH}$ in is considered to be arbitrary, and it is not sufficient to use Hartle-Hawking's boundary conditions to fix $D_{HH}$, thereby some additional conditions are required to determine it. To avoid the occurrence of imaginary temperature in a region outside the horizon and to prevent temperatures that are inversely related to distance, the temperature remains real throughout the entire region and decreases monotonically as $r$ increases, for $D_{HH} \geq D_c \simeq 23.03$, then reaching a maximum value at $r_c \simeq 1.43r_h$. In what follows, $D_{HH} \geq D_c$ is satisfied. Besides, it can be directly seen that the local temperature vanishes at the horizon $(r = r_h)$, and approaches the expected Hawking temperature as $r$ tends to infinity $(r \to \infty)$\cite{kaczmarek2024signatures,kaczmarek2026nonlocal}. By substituting the global Hawking temperature with the position-dependent local temperature $T_{HH}$ from Eq. \eqref{eq:THH}, we further explore and plot the evolutionary behavior of the relative-entropy imaginarity $\mathcal{I}_r$, as shown in FIG. \ref{fig:Ir_all_3D}.

Firstly, panels (a)-(c) depict the joint evolution of relative-entropy imaginarity for three distinct bipartite reduced states as functions of the normalized radial distance \(r/r_h\) and relative phase \(\phi\). For the physically accessible state \(\rho_{A_{\mathrm{I}}B_{\mathrm{I}}}\), the imaginarity exhibits unique dual-dimensional modulation characteristics. Along the phase dimension, it presents a symmetric double-valley structure. In the radial dimension, the imaginarity displays a regular single-valley profile: it gradually decays within the regime of the black hole quantum atmosphere and monotonically recovers toward a stable finite value at larger radial distances outside the atmospheric region. In comparison, the physically inaccessible cross-region state \(\rho_{A_{\mathrm{I}}B_{\mathrm{II}}}=\rho_{A_{\mathrm{II}}B_{\mathrm{I}}}\) and fully nonlocal state \(\rho_{A_{\mathrm{II}}B_{\mathrm{II}}}\) share qualitatively identical structural features. Both inaccessible states exhibit symmetric double-peak modulation with respect to the relative phase \(\phi\) and follow a single-peak evolutionary trend along the radial direction. Notably, the amplitude of imaginarity decreases hierarchically with increasing spacetime nonlocality: the cross-region state possesses a moderate imaginary magnitude, while the imaginarity of the fully nonlocal state \(\rho_{A_{\mathrm{II}}B_{\mathrm{II}}}\) is drastically suppressed to the order of \(10^{-3}\).

Further, panels (d)-(f) illustrate the pure radial variation of imaginarity under different Hartle--Hawking parameters \(D_{HH}\), systematically verifying the structural properties observed in the three-dimensional profiles. For the physically accessible state \(\rho_{A_{\mathrm{I}}B_{\mathrm{I}}}\), the radial imaginarity consistently decreases first and then increases with growing \(r/r_h\), forming a stable minimum value within the black hole quantum atmosphere region. For the physically inaccessible states \(\rho_{A_{\mathrm{I}}B_{\mathrm{II}}}\) and \(\rho_{A_{\mathrm{II}}B_{\mathrm{II}}}\), the radial imaginarity exhibits an opposite trend, rising first and then falling to form a prominent peak outside the event horizon. Moreover, the fully nonlocal state \(\rho_{A_{\mathrm{II}}B_{\mathrm{II}}}\) possesses a far weaker peak amplitude than the cross-region state \(\rho_{A_{\mathrm{I}}B_{\mathrm{II}}}\), revealing that the nonlocal gravitational environment induced by curved spacetime significantly depletes imaginary quantum resources.

The radial position of the imaginarity extremum is universally governed by the parameter \(D_{HH}\). For each fixed \(D_{HH}\), the extremal positions of imaginarity for all three quantum states coincide exactly, indicating that the radial response of imaginary quantum resources is uniformly determined by the spatially dependent Hartle--Hawking local temperature distribution. As \(D_{HH}\) increases, the extremum shifts toward larger \(r/r_h\), yet it always remains confined within the range \(1.4325 \lesssim r/r_h<1.5\). This interval matches the extremal region of local Hawking radiation in the black hole quantum atmosphere, demonstrating that the relative-entropy imaginarity of bipartite reduced states can effectively track the radial peak feature of Hawking radiation. 

\subsection*{B. Geometric imaginarity }
The above analyses demonstrate that the relative-entropy imaginarity \(\mathcal{I}_r\) can characterize the imaginary quantum resources induced by black hole spacetime, and its extremal evolution behaviors precisely capture the radial features of near-horizon Hawking radiation. Despite its rigorous thermodynamic physical interpretation, the calculation of \(\mathcal{I}_r\) for generic mixed states requires cumbersome spectral decomposition and logarithmic operations, which restricts its numerical feasibility in complex gravitational quantum systems. To remedy these limitations and achieve numerically efficient quantification of quantum imaginarity, we introduce the geometric imaginarity \(\mathcal{I}_g\). This alternative measure quantifies imaginary quantum features by evaluating the geometric deviation between a given density matrix and its complex conjugate counterpart on the quantum state manifold, providing a complementary geometric perspective to the thermodynamic description of \(\mathcal{I}_r\). The corresponding explicit definition is presented as follows:

\begin{equation}
\mathcal{I}_{g}(\rho)=\frac{1}{2}\bigl(1-\sqrt{F\bigl(\rho, \rho^{*}\bigr)}\bigr),
\label{eq:Ig_def}
\end{equation}
where \(\rho^{*}\) denotes the complex conjugate state of \(\rho\), and $F(\rho,\rho^{*})=\bigl(\operatorname{tr}\sqrt{\rho^{1/2}\rho^{*}\rho^{1/2}}\bigr)^{2}$ is the Uhlmann fidelity\cite{wu2021resource}. Different from entropy-based imaginarity, \(\mathcal{I}_g\) directly connects quantum imaginarity to the geometric distance defined on the state, avoiding complicated spectral computations and enabling efficient numerical quantification of imaginary resources for quantum states in curved spacetime.

Substituting the reduced states of the accessible and inaccessible bipartite subsystems into Eq. \eqref {eq:Ig_def}, we arrive at the analytical geometric imaginarity expressions listed below:
\begin{align}
	\mathcal{I}_g(\rho_{A_{1}B_{1}})
	&= \frac{1}{2}-\sin^2 r \cos^2 r \cos^2\frac{\theta}{2}
	- \frac{\sqrt{2}}{8}\sqrt{q_{11}},
	\label{eq:Ig_A1B1} \\[8pt]
	\begin{split}
	\mathcal{I}_g(\rho_{A_{1}B_{2}})&=\mathcal{I}_g(\rho_{A_{2}B_{1}})
	= \frac{1}{2}-\frac{1}{2}\sin^4 r \cos^2\frac{\theta}{2}\\ 
    &\quad- \frac{1}{2}\cos^4 r \cos^2\frac{\theta}{2}
	- \frac{1}{2}\sqrt{q_{12}},
	\end{split}
	\label{eq:Ig_cross_pair} \\[8pt]
	\mathcal{I}_g(\rho_{A_{2}B_{2}})
	&= \frac{1}{2}-\sin^2 r \cos^2 r \cos^2\frac{\theta}{2}
	- \frac{\sqrt{2}}{8}\sqrt{q_{22}}.
	\label{eq:Ig_A2B2}
\end{align}
where parameters are provided,
\begin{align*}
	q_{11}
	&= 8\left(\sin^4 r \cos^2\frac{\theta}{2} + \sin^2\frac{\theta}{2}\right)^{\!2} \\
	&\quad + 2(1-\cos2\theta)\cos^4 r \cos2\phi \\
	&\quad + 16\sin^4 r \cos^4 r \cos^4\frac{\theta}{2} 
	+ 8\cos^8 r \cos^4\frac{\theta}{2}, \\[6pt]
	q_{12}
	&= \left(\sin^2 r \cos^2 r \cos^2\frac{\theta}{2} + \sin^2\frac{\theta}{2}\right)^{\!2} \\
	&\quad + 2\sin^2 r \cos^2 r \sin^2\frac{\theta}{2} \cos^2\frac{\theta}{2}\cos2\phi \\
	&\quad + 3\sin^4 r \cos^4 r \cos^4\frac{\theta}{2}, \\[6pt]
	q_{22}
	&= 8\left(\sin^2\frac{\theta}{2} + \cos^4 r \cos^2\frac{\theta}{2}\right)^{\!2} \\
	&\quad + 2(1-\cos2\theta)\sin^4 r \cos2\phi \\
	&\quad + 8\sin^8 r \cos^4\frac{\theta}{2}
	+ 16\sin^4 r \cos^4 r \cos^4\frac{\theta}{2}.
\end{align*}

	As we have already introduced the radial-dependent local temperature $T_{HH}$ of the Hartle–Hawking vacuum in the preceding sections, we adopt an identical treatment here by substituting the standard global Hawking temperature $T$ with $T_{HH}$ from Eq. \eqref{eq:THH}. With this setup, we compute and visualize the geometric imaginarity quantities for all bipartite reduced states in FIG. 2.

    Firstly, panels (a)--(c) in Fig. 2 illustrate the geometric imaginarity of three distinct bipartite reduced states as functions of angular variables $\theta$ and $\phi$ at fixed radial coordinate $r/r_h=1.5$. For all three configurations, the geometric imaginarity exhibits a symmetric bimodal profile with respect to $\phi=\pi$, while the imaginarity vanishes at the boundaries $\phi=0$ and $\phi=2\pi$. The polar angle $\theta$ modulates the height of the double-peak structure without altering the intrinsic symmetry. Hierarchical suppression of amplitude arises with growing causal isolation: the physically accessible state $\rho_{A_{\mathrm{I}}B_{\mathrm{I}}}$ sustains the largest imaginary magnitude, followed by the cross-region state $\rho_{A_{\mathrm{I}}B_{\mathrm{II}}}=\rho_{A_{\mathrm{II}}B_{\mathrm{I}}}$, whereas the imaginarity of the fully inaccessible state $\rho_{A_{\mathrm{II}}B_{\mathrm{II}}}$ is heavily suppressed down to the order of $10^{-3}$.
    
    For the (d)--(f) in Fig. 2, the geometrical imaginarity for the three bipartite reduced states exhibits distinct radial behaviors against $r/r_h$. Specifically, the geometrical imaginarity of the physically accessible state $\rho_{A_{\mathrm{I}}B_{\mathrm{I}}}$ first decreases and then increases with growing radial distance $r/r_h$, forming a clear local minimum within the quantum atmosphere and gradually approaching a saturated value in the far region. In contrast, the geometrical imaginarity of the inaccessible configurations rises to a prominent local maximum and slowly decays afterward, which reveals that the local Hawking effect redistributes imaginary resources between accessible and inaccessible regions.

\begin{figure*}[tb]
	\centering
	\subfloat[]{%
		\includegraphics[width=0.32\textwidth]{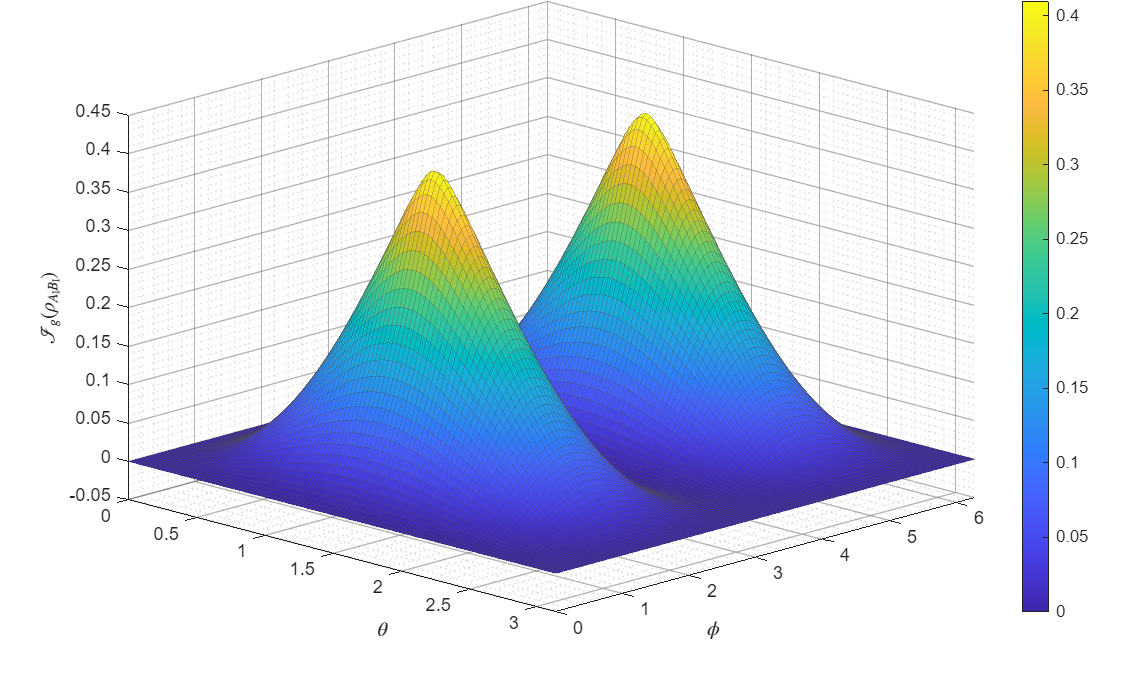}%
		\label{subfig:Ig_AIBI_curve}%
	}
	\hfill
	\subfloat[]{%
		\includegraphics[width=0.32\textwidth]{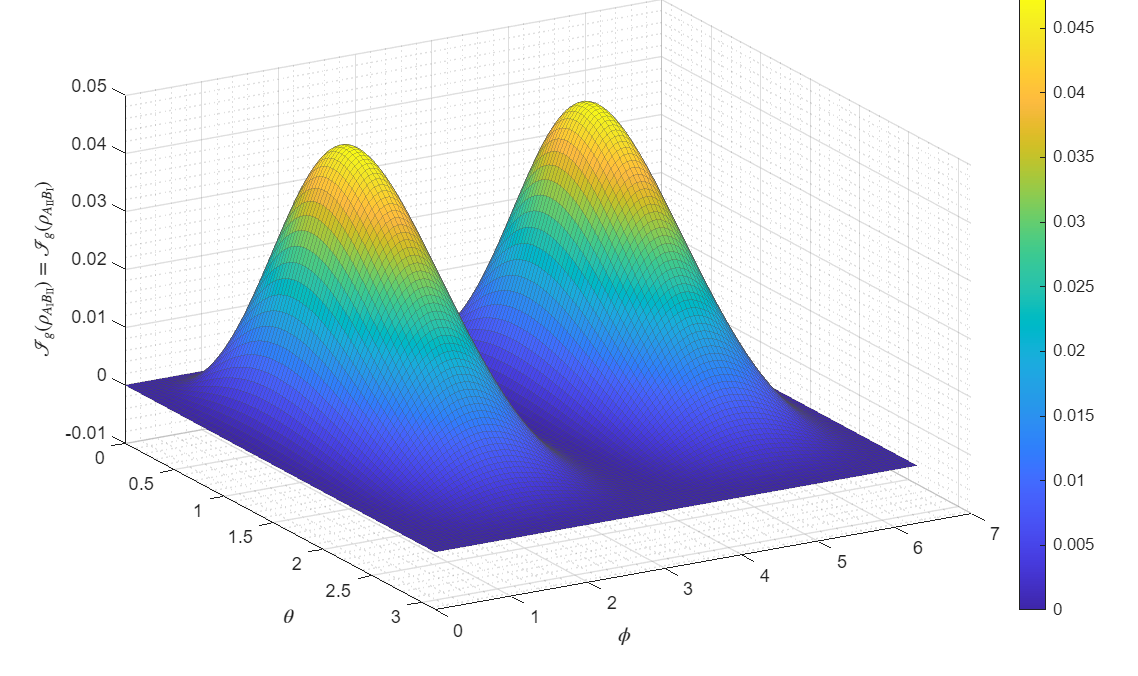}%
		\label{subfig:Ig_AIBII_curve}%
	}
	\hfill
	\subfloat[]{%
		\includegraphics[width=0.32\textwidth]{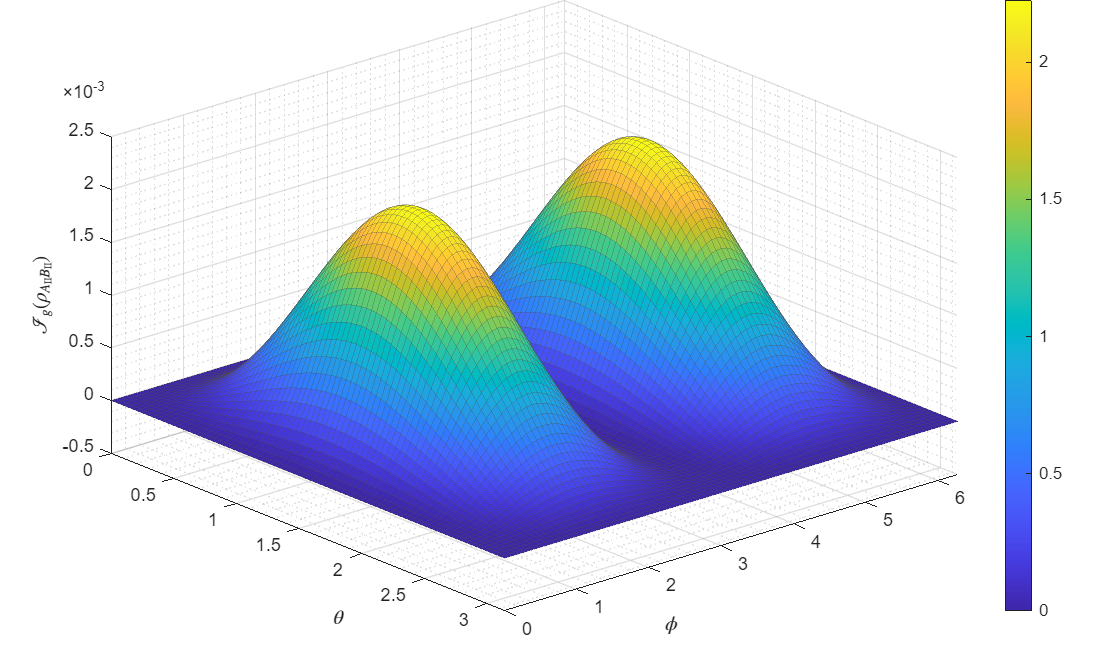}%
		\label{subfig:Ig_AIIBII_curve}%
	}

	\vspace{0.6em}

	\subfloat[]{%
		\includegraphics[width=0.32\textwidth]{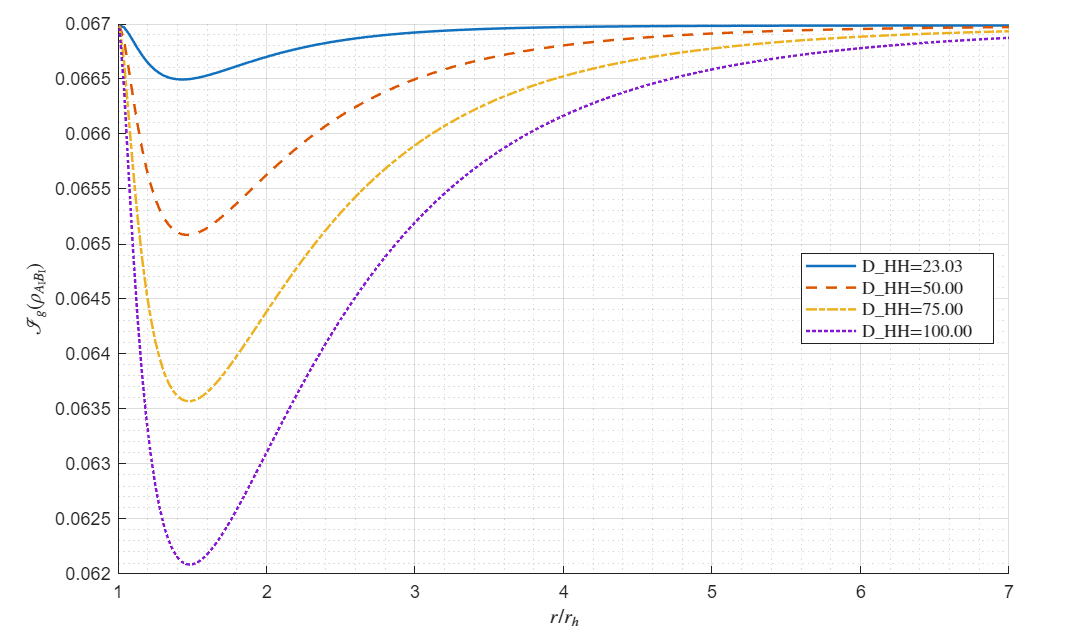}%
		\label{subfig:Ig_AIBI_surf}%
	}
	\hfill
	\subfloat[]{%
		\includegraphics[width=0.32\textwidth]{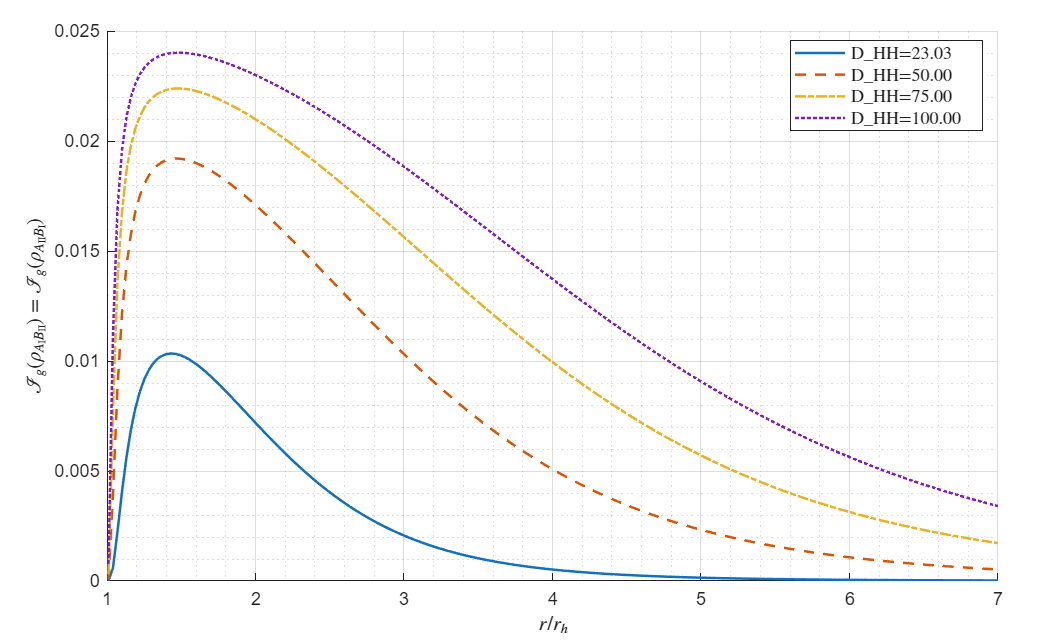}%
		\label{subfig:Ig_AIBII_surf}%
	}
	\hfill
	\subfloat[]{%
		\includegraphics[width=0.32\textwidth]{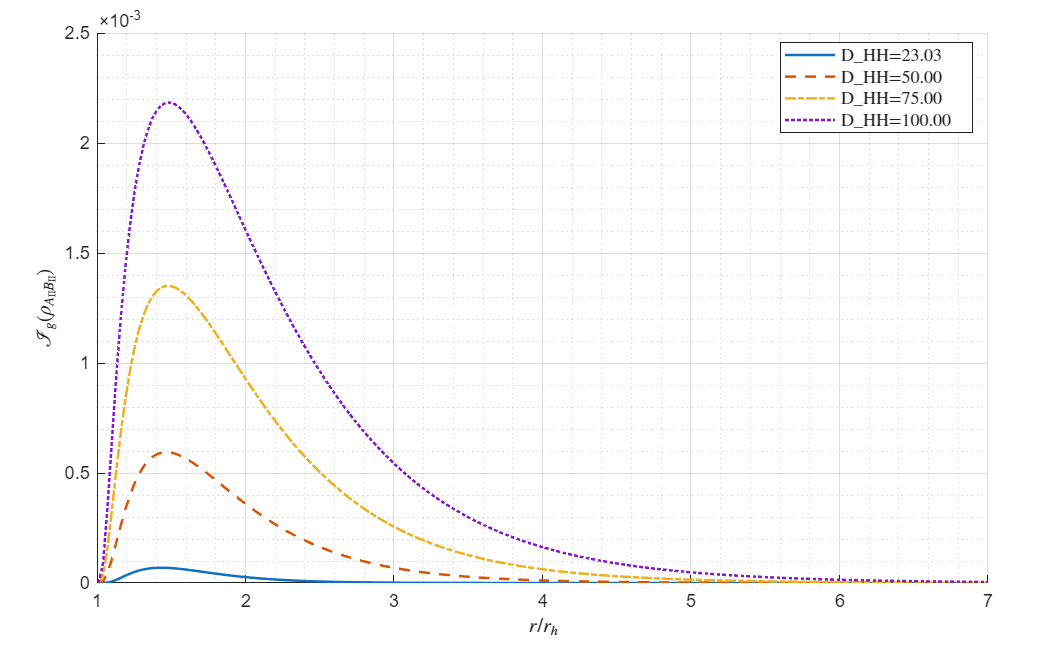}%
		\label{subfig:Ig_AIIBII_surf}%
	}

	\caption{The geometric imaginarity of the states $\rho_{A_{\mathrm{I}}B_{\mathrm{I}}}$, $\rho_{A_{\mathrm{I}}B_{\mathrm{II}}}$ and $\rho_{A_{\mathrm{II}}B_{\mathrm{II}}}$. Panels (a)--(c) are three‑dimensional surface plots as functions of the state parameter $\theta$ and the phase $\phi$, with normalized radial distance fixed at $r/r_h=1.5$, $D_{HH}=50$ and $\omega=1$. Panels (d)--(f) present the corresponding radial cross‑section curves varying with $r/r_h$ under fixed $\theta$ and $\phi$.}
	\label{fig:Ig_radial_theta}
\end{figure*}

\subsection*{C. Robustness-based imaginarity }

While $\mathcal{I}_g$ offers a transparent geometric characterization of quantum imaginarity built upon Uhlmann fidelity, this distance-based measure lacks a direct operational interpretation concerning how imaginary quantum resources endure environmental noise.

To fill this gap, we adopt the robustness of imaginarity $\mathcal{R}(\rho)$ to quantify the noise tolerance of imaginarity within the resource-theoretic framework \cite{xue2021quantificationimaginarity}. By definition,
\begin{equation}
\mathcal{R}(\rho)=\min_{\tau\in\mathcal{D}(\mathcal{H})}\left\{s\ge0\;\bigg|\; \frac{\rho+s\tau}{1+s}\in\mathcal{R}\right\},
\label{eq:RoI_def_general}
\end{equation}
where $\mathcal{D}(\mathcal{H})$ denotes the set of all density operators and $\mathcal{R}$ represents the set of real quantum states. This quantity describes the minimal weight of auxiliary quantum states mixed with $\rho$ such that the resulting state contains no imaginary components. For the convenience of evaluation, the robustness of imaginarity can be recast into an equivalent  expression in terms of the trace norm:
\begin{equation}
\begin{aligned}
\mathcal{R}(\rho)
&= \frac{1}{2}\left\lVert \rho - \rho^\mathsf{T} \right\rVert_1.
\end{aligned}
\label{eq:RoI_def}
\end{equation}

By employing Eq. \eqref{eq:RoI_def}, we derive the robustness-based imaginarity of the above reduced bipartite states :

\begin{align}
	\mathcal{R}(\rho_{A_{\mathrm{I}}B_{\mathrm{I}}})
	&= \cos^2 r \left| \sin\theta \sin\phi \right|,
	\label{eq:RoI_A1B1} \\[4pt]
	\mathcal{R}(\rho_{A_{\mathrm{I}}B_{\mathrm{II}}})
	&=\mathcal{R}(\rho_{A_{\mathrm{II}}B_{\mathrm{I}}})= \frac{1}{2}\left| \sin2r\sin\theta \sin\phi \right|,
	\label{eq:RoI_cross} \\[4pt]
	\mathcal{R}(\rho_{A_{\mathrm{II}}B_{\mathrm{II}}})
	&= \sin^2 r\left| \sin\theta \sin\phi \right|.
	\label{eq:RoI_A2B2}
\end{align}

It is interesting to note a unified quantitative relation connects all three quantities. We can rewrite the cross term as
\begin{equation}
	\mathcal{R}(\rho_{A_{\mathrm{I}}B_{\mathrm{II}}})
	=\sqrt{\mathcal{R}(\rho_{A_{\mathrm{I}}B_{\mathrm{I}}})\mathcal{R}(\rho_{A_{\mathrm{II}}B_{\mathrm{II}}})}.
\end{equation}

Thus, FIG. 3 demonstrates that the local Hawking effect triggers a redistribution of imaginary resources between physically accessible and inaccessible bipartite configurations, where the robustness of imaginarity for all three bipartite partitions exhibits evident angular and radial dependence with fixed $\phi=\pi/4$ and $\omega=1$. For every state, the robustness of imaginarity rises and then falls with varying $\theta$, reaching its maximum at $\theta=\pi/2$. Meanwhile, distinct radial behaviors emerge for different observer partitions. The physically accessible state $\rho_{A_{\mathrm{I}}B_{\mathrm{I}}}$ exhibits valley-shaped variation against $r/r_h$, with a well-defined local minimum emerging within the quantum atmospheric zone before gradually converging to a saturated plateau at large radii. In contrast, the partially inaccessible state $\rho_{A_{\mathrm{I}}B_{\mathrm{II}}}$ and fully inaccessible state $\rho_{A_{\mathrm{II}}B_{\mathrm{II}}}$ both develop prominent local maxima as $r/r_h$ increases from the horizon outward.

These radial features are further verified by the radial profiles plotted in (d)-(f). The extremal radial location relies on the Hartle--Hawking parameter $D_{HH}$. For a fixed $D_{HH}$, the extrema of the robustness of imaginarity for all three states appear at nearly identical radial positions, demonstrating that their radial response is controlled by the same local Hawking thermal distribution. As $D_{HH}$ grows, these extremal positions shift toward larger $r/r_h$, yet are always confined within $1.4322 \lesssim r/r_h<1.5$. This interval coincides with the extremal region of local Hawking radiation inside the quantum atmosphere, showing that the robustness of imaginarity of bipartite reduced states traces the radial peak of Hawking radiation.

\begin{figure*}[tb]
	\centering
	\subfloat[]{%
		\includegraphics[width=0.32\textwidth]{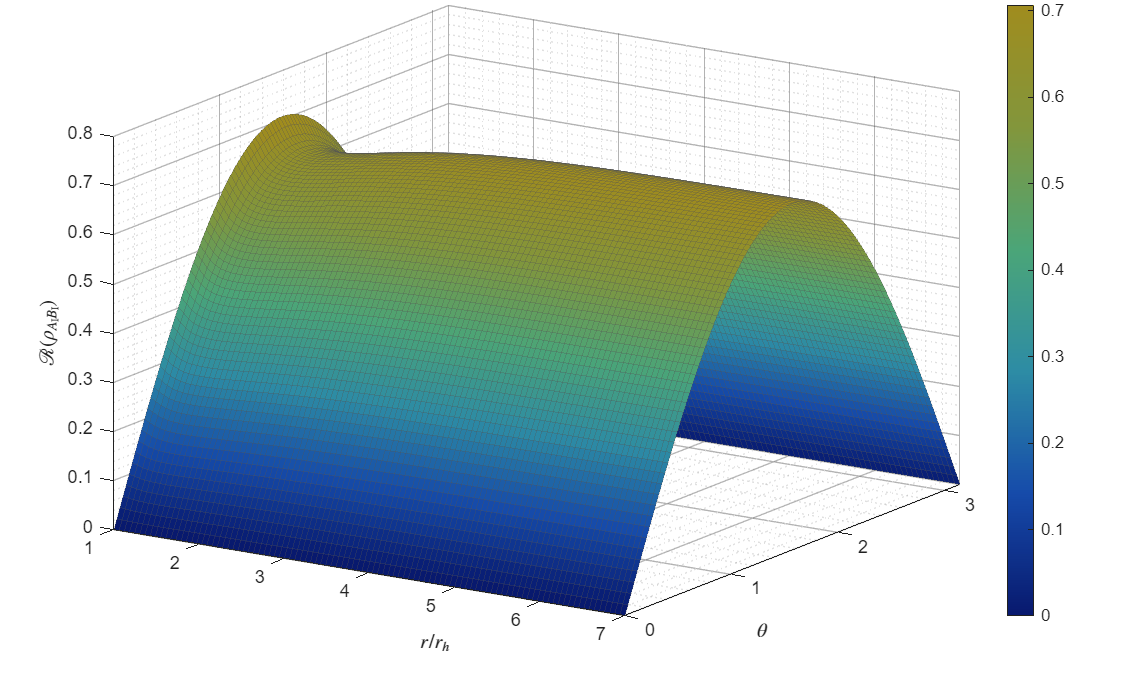}%
		\label{subfig:RoI_A1B1_curve}%
	}
	\hfill
	\subfloat[]{%
		\includegraphics[width=0.32\textwidth]{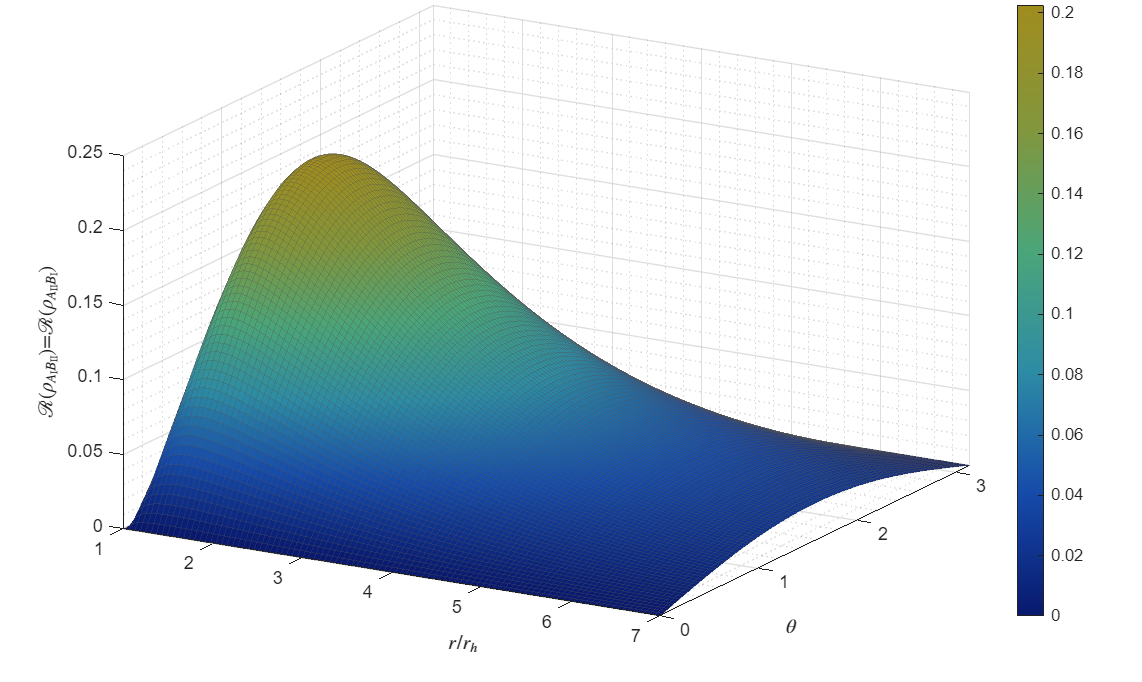}%
		\label{subfig:RoI_A1B2_curve}%
	}
	\hfill
	\subfloat[]{%
		\includegraphics[width=0.32\textwidth]{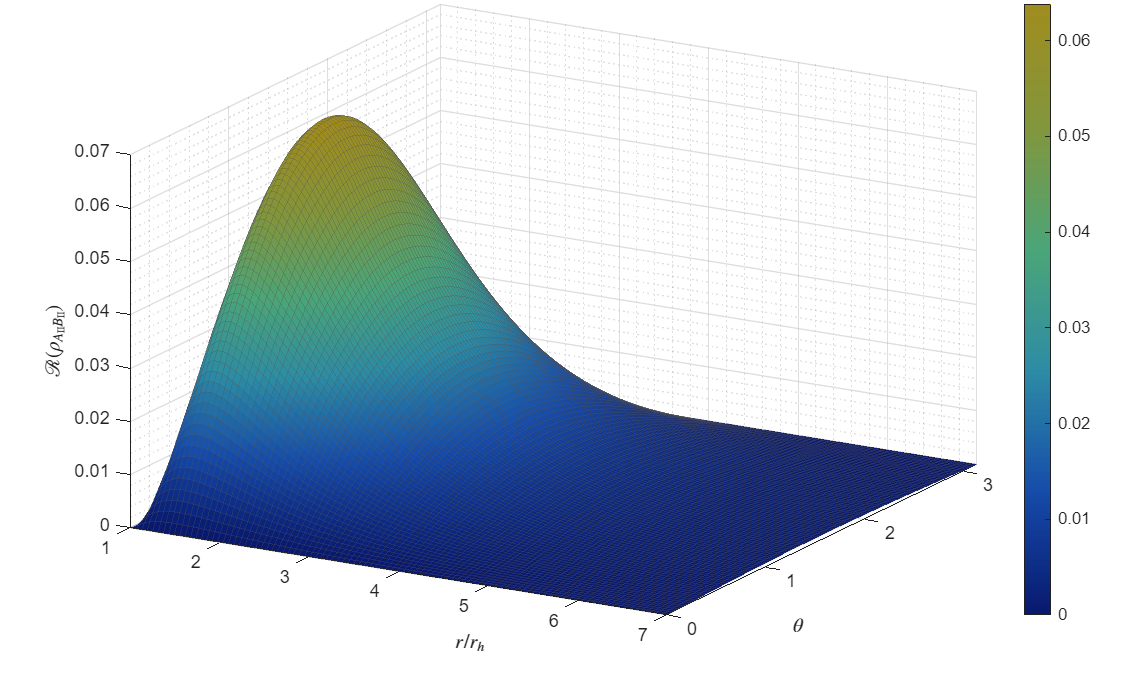}%
		\label{subfig:RoI_A2B2_curve}%
	}

	\vspace{0.6em}

	\subfloat[]{%
		\includegraphics[width=0.32\textwidth]{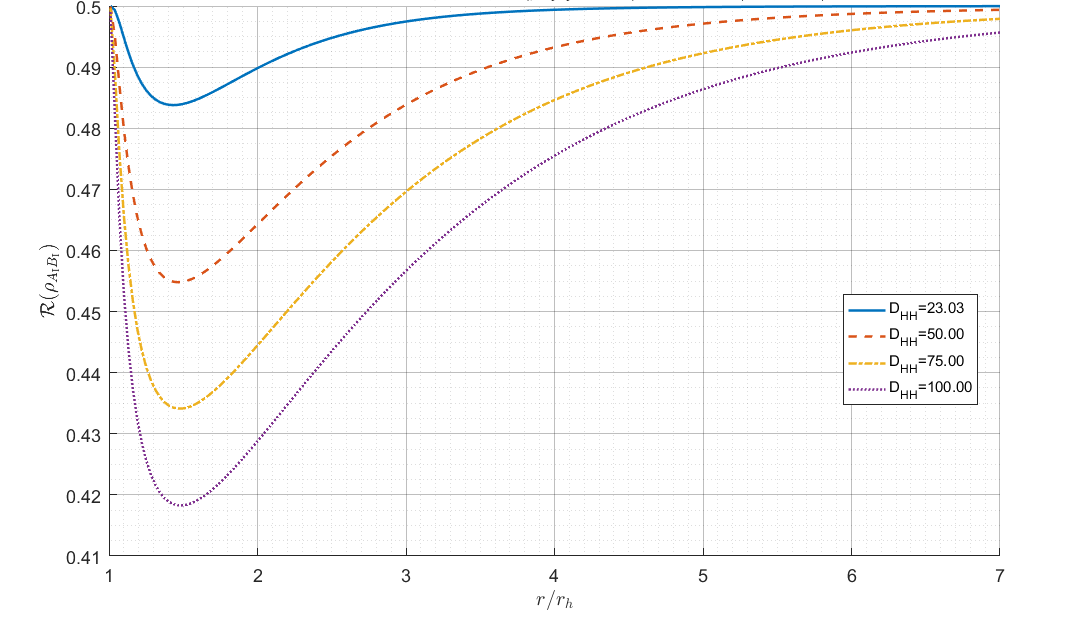}%
		\label{subfig:RoI_A1B1_surf}%
	}
	\hfill
	\subfloat[]{%
		\includegraphics[width=0.32\textwidth]{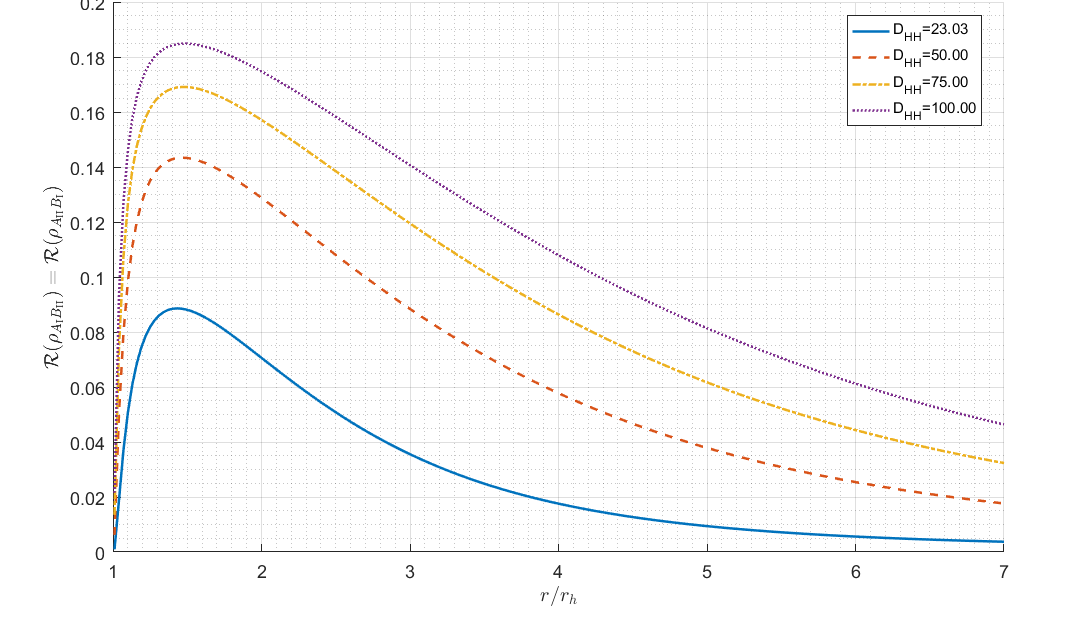}%
		\label{subfig:RoI_A1B2_surf}%
	}
	\hfill
	\subfloat[]{%
		\includegraphics[width=0.32\textwidth]{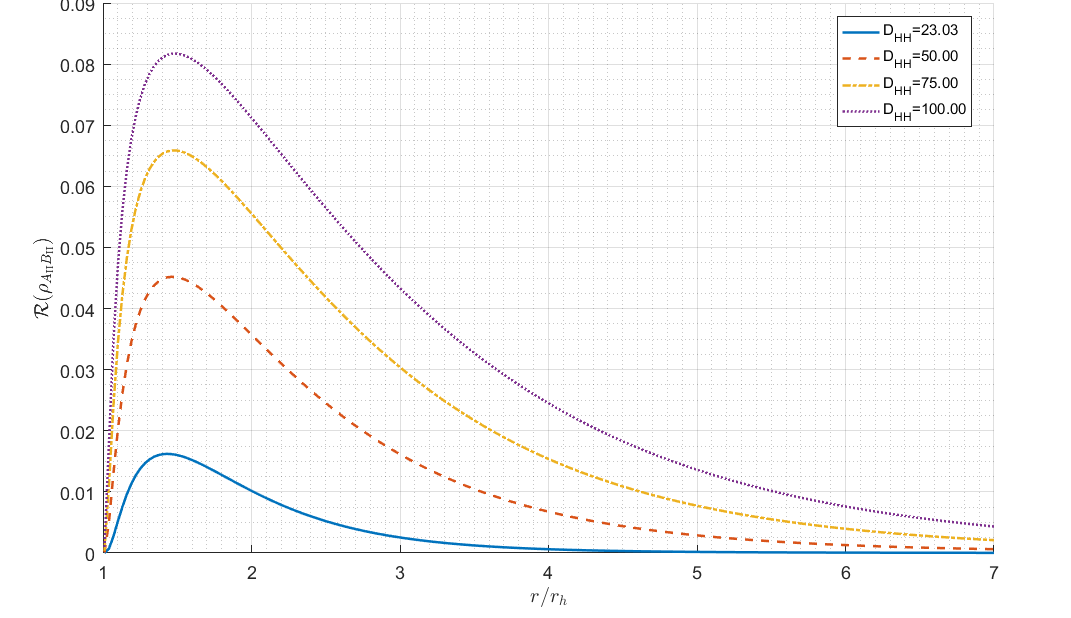}%
		\label{subfig:RoI_A2B2_surf}%
	}

	\caption{The robustness of imaginarity for the states $\rho_{A_{\mathrm{I}}B_{\mathrm{I}}}$, $\rho_{A_{\mathrm{I}}B_{\mathrm{II}}}$ and $\rho_{A_{\mathrm{II}}B_{\mathrm{II}}}$ as functions of the normalized radial distance $r/r_h$ and the state parameter $\theta$, with fixed parameters $D_{HH}=50$, $\phi=\pi/4$ and $\omega=1$. Panels (a)--(c) correspond to three‑dimensional surface plots, and panels (d)--(f) display cross‑section curves under different values of $D_{HH}$.}
	\label{fig:RoI_radial_theta}
\end{figure*}

To quantitatively characterize the radial behavior of the quantum imaginarity, we summarize their extremal positions in TABLE~\ref{tab:imaginarity_extrema}. As $D_{HH}$ increases from 23.03 to 100, the extrema shift outward from $r_{\rm ext}/r_{h}=1.4322$ to approximately 1.48, while remaining confined within the finite near-horizon interval $1.4322 \lesssim r/r_{h}<1.5$. Their extremal positions nearly coincide, suggesting that relative-entropy imaginarity, geometric imaginarity, and robustness of imaginarity are all sensitive to the local Hawking effect in the same radial region. However, these quantities characterize distinct aspects: relative-entropy imaginarity originates from the relative entropy distance and reveals the energetic feature of imaginary resources; geometric imaginarity describes quantum imaginarity from a fidelity-based geometric perspective; and robustness of imaginarity quantifies the noise tolerance of imaginary resources within the resource-theoretic framework. Thus, they serve as complementary signatures of the local Hawking effect, with their extrema consistently tracking the same near-horizon atmospheric region.

\begin{table}[htb]
	\begin{ruledtabular}
		\begin{tabular}{cccc}
			\multicolumn{4}{c}{Extremal radial position $r/r_h$} \\
			\hline
			$D_{HH}$ & $\mathcal{I}_r$ & $\mathcal{I}_g$ & $\mathcal{R}$ \\
			\hline
			23.03  & 1.43255 & 1.43256 & 1.43256 \\
			50.00  & 1.46650 & 1.46650 & 1.46650 \\
			75.00  & 1.47720 & 1.47720 & 1.47720 \\
			100.00 & 1.48271 & 1.48272 & 1.48272 \\
		\end{tabular}
	\end{ruledtabular}
	\captionsetup{justification=raggedright,singlelinecheck=false}
	\caption{\label{tab:imaginarity_extrema} Extremal radial positions $r/r_h$ of $\mathcal{I}_r$, $\mathcal{I}_g$, and $\mathcal{R}$ for different $D_{HH}$. The values correspond to the extrema of the physically accessible and inaccessible bipartite states.}

\end{table}

\section{CONCLUSIONS AND DISCUSSIONS}
In this work, based on the four-partite Dirac field state in a static Schwarzschild black hole spacetime, we systematically investigate the influence of Hawking thermal radiation within the black hole quantum atmosphere on three canonical quantum imaginarity measures of bipartite reduced states, i.e., relative-entropy imaginarity, geometric imaginarity, and robustness of imaginarity.
Numerical simulations adopt distinct variable combinations for each metric. For relative-entropy imaginarity, we explore its joint dependence on normalized radial distance $r/r_h$ and relative phase $\phi$. For geometric imaginarity, we fix $r/r_h=1.5$ and examine angular modulation governed by $\theta$ and $\phi$. For the robustness of imaginarity, we analyze its evolution against $r/r_h$ and polar angle $\theta$ at fixed $\phi=\pi/4$.

Specifically, the fully accessible state $\rho_{A_{\mathrm{I}}B_{\mathrm{I}}}$ exhibits a universal valley-shaped profile as $r/r_h$ grows, while the cross-coupling state and fully inaccessible state follow opposite peak-shaped evolutionary trends. For relative-entropy imaginarity, the accessible configuration presents a symmetric double-valley structure in the phase dimension, whereas cross and fully inaccessible states show symmetric double-peak patterns; the imaginarity magnitude is hierarchically suppressed with increasing causal isolation, falling to the order of $10^{-3}$ for $\rho_{A_{\mathrm{II}}B_{\mathrm{II}}}$. At fixed radial coordinate, geometric imaginarity for all partitions features symmetric bimodal distributions with respect to $\phi=\pi$ and vanishes at $\phi=0,2\pi$, with the polar angle only modulating peak heights without breaking symmetry. As for imaginarity robustness, all states reach maximum values at $\theta=\pi/2$.
Remarkably, all three imaginarity measures display consistent extremum signatures near $r/r_h\approx 1.43$, which closely correlates with the peak intensity of Hawking thermal radiation inside the quantum atmosphere. Further analysis on model parameters reveals that the Hartle–Hawking constant $D_{HH}$ markedly amplifies the redistribution of imaginary quantum resources between physically accessible and inaccessible regions. Increasing the horizon radius $r_h$, by contrast, weakens such redistribution and smoothes out the extremum features within the quantum atmospheric zone.

These findings provide novel insights into the information structure of black hole quantum atmospheres and the intrinsic quantum nature of Hawking radiation from the perspective of imaginary quantum resources.
\section*{ACKNOWLEDGMENTS}

This work is supported by the Natural Science Foundation of Hainan Province under Grant No. 125RC744; the China Scholarship Council (CSC).
\nocite{*}
\bibliography{references}

\appendix
\renewcommand{\thesection}{}       
\section{}
\label{app:aux_params}
 Due to the Hawking radiation of the black hole, the Dirac field becomes modified from the perspective of a uniformly moving observer. So the state $\rho_{AB}$ will be transformed to a four-partite quantum state $\rho_{A_{\mathrm{I}}A_{\mathrm{II}}B_{\mathrm{I}}B_{\mathrm{II}}}$.
By tracing over the modes $A_{\mathrm{II}}$ and $B_{\mathrm{II}}$, we obtain the reduced bipartite mixed states 
\begin{equation}
\begin{split}
	&\rho_{A_{\mathrm{I}}B_{\mathrm{I}}}=\\
	&\begin{pmatrix}
		c^2(1-t)^2 & 0 & 0 & e^{-i\phi}cs(1-t) \\
		0 & c^2 t(1-t) & 0 & 0 \\
		0 & 0 & c^2 t(1-t) & 0 \\
		e^{i\phi}cs(1-t) & 0 & 0 & c^2 t^2 + s^2
	\end{pmatrix},
    \end{split}
\end{equation}
\text{where } $c=\cos\tfrac{\theta}{2},\ s=\sin\tfrac{\theta}{2},\ t=\sin^2 r$.

By tracing over the modes $A_{\mathrm{II}}$ and $B_{\mathrm{I}}$, we obtain the reduced bipartite mixed states 
\begin{equation}
\begin{split}
	&\rho_{A_{\mathrm{I}}B_{\mathrm{II}}}=\\
	&\begin{pmatrix}
		c^2(1-t)^2 & 0 & 0 & 0 \\
		0 & c^2 t(1-t) & e^{-i\phi} cs\sqrt{t(1-t)} & 0 \\
		0 & e^{i\phi} cs\sqrt{t(1-t)} & c^2 t(1-t) + s^2 & 0 \\
		0 & 0 & 0 & c^2 t
	\end{pmatrix},
	\label{eq:rho_A1B2_short}
    \end{split}
\end{equation}
\text{where } $c=\cos\frac{\theta}{2},\quad s=\sin\frac{\theta}{2},\quad t=\sin^2 r$.

By tracing over the modes $A_{\mathrm{I}}$ and $B_{\mathrm{II}}$, we obtain the reduced bipartite mixed states $\rho_{A_{\mathrm{II}}B_{\mathrm{I}}}$,

\begin{equation}
\begin{split}
		&\rho_{A_{\mathrm{II}}B_{\mathrm{I}}}=\\
		&\begin{pmatrix}
			c^2(1-t)^2 & 0 & 0 & 0 \\
			0 & c^2t(1-t)+s^2 & e^{i\phi}cs\sqrt{t(1-t)} & 0 \\
			0 & e^{-i\phi}cs\sqrt{t(1-t)} & c^2t(1-t) & 0 \\
			0 & 0 & 0 & c^2t^2
		\end{pmatrix},
	\label{eq:rho_A2B1_slim}
    \end{split}
\end{equation}
\text{where } $c=\cos\tfrac{\theta}{2},\ s=\sin\tfrac{\theta}{2},\ t=\sin^2 r$.

By tracing over the modes $A_{\mathrm{I}}$ and $B_{\mathrm{I}}$, we obtain the reduced bipartite mixed states 
\begin{equation}
	\begin{split}
		&\rho_{A_{\mathrm{II}}B_{\mathrm{II}}}=\\
		&\begin{pmatrix}
			c^2(1-t)^2+s^2 & 0 & 0 & e^{i\phi}cs\,t \\
			0 & c^2 t(1-t) & 0 & 0 \\
			0 & 0 & c^2 t(1-t) & 0 \\
			e^{-i\phi}cs\,t & 0 & 0 & c^2 t^2
		\end{pmatrix},
	\label{eq:rho_A2B2_final_correct}
    \end{split}
\end{equation}
\text{where } $c=\cos\tfrac{\theta}{2},\ s=\sin\tfrac{\theta}{2},\ t=\sin^2 r$.

Thus, the relative-entropy imaginarity of these states can be derived from Eq. \eqref{eq:Ir_def},

\begin{align}
\begin{split}
	\mathcal{I}_r(\rho_{A_{\mathrm{I}}B_{\mathrm{I}}})
	&=\frac{1}{2\log 2}
	\Bigl[
	f(u_{11},v_{11},w_{11})
	-f(\overline{u}_{11},v_{11},w_{11})
	\Bigr],
	\\[6pt]
    \end{split}
\end{align}
\begin{align}
\begin{split}
	\mathcal{I}_r(\rho_{A_{\mathrm{I}}B_{\mathrm{II}}})
    &=\mathcal{I}_r(\rho_{A_{\mathrm{II}}B_{\mathrm{I}}})\\
	&=\frac{1}{2\log 2}
	\Bigl[
	f(u_{12},v_{12},w_{12})
	-f(\overline{u}_{12},v_{12},w_{12})
	\Bigr],
	\label{eq:Ir_cross}\\[6pt]
     \end{split}
\end{align}
\begin{align}
\begin{split}
	\mathcal{I}_r(\rho_{A_{\mathrm{II}}B_{\mathrm{II}}})
	&=\frac{1}{2\log 2}
	\Bigl[
	f(u_{22},v_{22},w_{22})
    -f(\overline{u}_{22},v_{22},w_{22})
	\Bigr],
	\label{eq:Ir_A2B2}
      \end{split}
\end{align}
where the function $f(u,v,w)$ is defined as
\begin{equation}
	\begin{split}
		f(u,v,w)
		&=\bigl(\sqrt{u}+v\bigr)\log\!\left(w+\frac{\sqrt{u}}{2}\right) \\
		&\quad-\bigl(\sqrt{u}-v\bigr)\log\!\left(w-\frac{\sqrt{u}}{2}\right),
	\end{split}
\end{equation}
where the auxiliary parameters are provided,
\begin{align*}
	u_{11}&=4c_\theta^2c_r^4-4c_\theta^2c_r^2+1,\\
	\overline{u}_{11}&=(2c_\theta^2c_r^4+s_\theta^2)^2
	-4c_\theta^2c_r^4(s_\phi^2s_\theta^2-s_r^4s_\theta^2+s_r^4),\\
	v_{11}&=2c_\theta^2c_r^4+s_\theta^2,\\
	w_{11}&=c_\theta^2c_r^4+\tfrac{1}{2}s_\theta^2,\\[4pt]
	u_{12}&=4c_\theta^2s_r^2c_r^2+s_\theta^2,\\
	\overline{u}_{12}&=4c_\theta^2s_r^2c_r^2c_\phi^2+s_\theta^2,\\
	v_{12}&=2c_\theta^2s_r^2c_r^2+s_\theta^2,\\
	w_{12}&=c_\theta^2s_r^2c_r^2+\tfrac{1}{2}s_\theta^2,\\[4pt]
	u_{22}&=4c_\theta^2c_r^4-4c_\theta^2c_r^2+1,\\
	\overline{u}_{22}&=(c_\theta^2s_r^4+s_\theta^2+c_\theta^2c_r^4)^2
	-4(s_\theta^2-s_\theta^2c_\phi^2+c_\theta^2c_r^4)c_\theta^2s_r^4,\\
	v_{22}&=c_\theta^2s_r^4+s_\theta^2+c_\theta^2c_r^4,\\
	w_{22}&=\tfrac{1}{2}(c_\theta^2s_r^4+s_\theta^2+c_\theta^2c_r^4),
\end{align*}
with $c_\theta=\cos(\theta/2)$, $s_\theta=\sin(\theta/2)$,
$c_r=\cos r$, $s_r=\sin r$, $c_\phi=\cos\phi$, and $s_\phi=\sin\phi$.
\end{document}